\documentclass[fleqn,twoside]{article}
\usepackage{espcrc2}
\usepackage{amssymb}

\usepackage{graphicx}
\usepackage{epsfig}
\newcommand{\beqn}{\begin{eqnarray}}
\newcommand{\eeqn}{\end{eqnarray}}
\newcommand{\beq}{\begin{equation}}
\newcommand{\eeq}{\end{equation}}
\newcommand{\eq}[1]{(\ref{#1})}

\newcommand{\cA}{{\cal A}}
\newcommand{\GeV}{~\mathrm{GeV}}

\newcommand{\offdiag}{\mathrm{offdiag}}
\newcommand{\diag}{\mathrm{diag}}

\title{
\thispagestyle{empty}
\vspace{-25mm}
\rightline{\small KANAZAWA-03-21~~~~~}
\rightline{\small ITEP-LAT-2003-14~~~~~}
\vspace{10mm}
Numerical study of gluon propagators
in Maximally Abelian gauge\thanks{Presented by S.~M.~M. at Lattice'03.}
}

\author{V.G.~Bornyakov\address{ITEP, B.Cheremushkinskaya 25, Moscow,117259,
Russia}${}^{,}$\address{Institute for High Energy Physics, Protvino 142284,
Russia}${}^{,}$\thanks{V.~G.~B. is partially supported by grants
RFBR 01-02-17456, INTAS-00-00111 and DFG-RFBR 436 RUS 113/739/0.},
M.N.~Chernodub${}^{\mathrm{a,}}$\address{Institute for Theoretical Physics, Kanazawa University,
Kanazawa 920-1192, Japan}${}^{,}$\thanks{M.~N.~Ch. is supported by JSPS Fellowship P01023.},
F.V.~Gubarev${}^{\mathrm{a,c,}}$\thanks{F.~V.~G. is supported by JSPS Fellowship P03024.},
S.M.~Morozov${}^{\mathrm{a,}}$\thanks{S.~M.~M. is partially supported by grant RFBR 02-02-17308
and CRDF award MO-011-0.},
M.I.~Polikarpov${}^{\mathrm{a,}}$\thanks{M.~I.~P is partially supported by grants RFBR
02-02-17308, RFBR 01-02-17456, RFBR 00-15-96-786, INTAS-00-00111,
and CRDF award RPI-2364-MO-02.}}

\begin{document}
\begin{abstract}
Propagators of diagonal and off--diagonal gluons are studied numerically
in the Maximally Abelian gauge of the SU(2) lattice gauge theory. We have found
the strong enhancement of the diagonal gluon in the infrared region.
The enhancement factor is about 50 at the smallest available momentum, 325~MeV.
We discuss also various analytical fits to the propagators.
\end{abstract}

\maketitle

Propagators of fundamental fields are basic quantities in quantum field theories.
In non--Abelian gauge theories the infrared behaviour of the gluon propagators
is believed to carry an information about the color confinement. A popular
approach to the confinement is based on the dual superconductor
mechanism~\cite{tHooftMandelstam} which is most thoroughly studied in the Maximally
Abelian (MA) gauge~\cite{ref:kronfeld} (see, {\it e.g.},~\cite{Review} for a review).
A study of the MA gauge gluon propagator in the coordinate space~\cite{Amemiya:zf}
has shown that the propagator of the off-diagonal gluons at large distances
is exponentially suppressed with respect to the diagonal propagator
by the effective mass about $1.2$~GeV.
This property of the gluon propagators -- supporting the Abelian dominance
in gluodynamics~\cite{Ezawa,Suzuki:1989gp} -- was confirmed and further explored
in the momentum space in Refs.~\cite{Bornyakov:2002vv,ref:itep:propagators}.
The mass gap generation for the off-diagonal gluons was also investigated
analytically~\cite{Offdiagonal:mass}. Below we provide numerical
results supporting the Abelian dominance in terms of the gluon propagators.

We study the SU(2) pure gauge model with the Wilson action.
We use the standard parameterization of $SU(2)$ link matrices
$U_{11}=\cos\varphi\, e^{i\theta}$ and $U_{12}=\sin\varphi\, e^{i\chi}$.
The gauge fields are given by
$$A^a_\mu(x) \,\sigma^a = - i (U_\mu(x) - U_\mu^{\dagger}(x))\,.$$
where $\sigma^a$ are the Pauli matrices.
In terms of the link angles one gets\footnote{Note that in Ref.~\cite{Bornyakov:2002vv}
the definition of the field $A$ differs from Eq.~\eq{eq:fields} by the factor of $2$.}:
\beqn
A^1_\mu(x) & = & 2 \sin\varphi_\mu(x)\sin\chi_\mu(x)\,,\nonumber\\
A^2_\mu(x) & = & 2 \sin\varphi_\mu(x)\cos\chi_\mu(x)\,,
\label{eq:fields} \\
A^3_\mu(x) & = & 2 \cos\varphi_\mu(x)\sin\theta_\mu(x)\,. \nonumber
\eeqn
We call $A_\mu^3(x)$ the diagonal gluon field, and
$A_\mu^i(x),\ i = 1,2$, the off-diagonal gluon field.

The MA gauge is fixed by maximizing
\beqn
F^{\mathrm{latt}}_{\mathrm{MAG}}[\varphi] = \sum\nolimits_{x,\mu}
\cos2\varphi_\mu(x)\,.
\label{eq:MAG:global}
\eeqn
This condition fixes the SU(2) gauge group up to a U(1) subgroup. In order to fix
the remaining U(1) gauge symmetry we implement a generalization of the
Landau gauge maximizing the functional
\beq
\label{eq:coscosfunc}
{\tilde F}^{\mathrm{latt}}_{\mathrm{Land}}[\theta, \varphi]=\sum\nolimits_{x,\mu}
\cos\varphi_\mu(x) \cos\theta_\mu(x)\,,
\eeq
which is consistent with the definition~\eq{eq:fields} of the $A^3_\mu$ field.
In the continuum  limit the definition \eq{eq:coscosfunc} corresponds to
the standard U(1) Landau gauge.

We calculate the diagonal and off-diagonal propagators,
\beqn
\label{eq:diagprop}
D_{\mu\nu}^{\mathrm{diag}}(p) & = &\langle\cA^3_\mu(k)\cA^3_\nu(-k)\rangle\,,\\
\label{eq:offdiagprop}
D_{\mu\nu}^{\mathrm{offdiag}}(p) & = &
\langle\cA^{+}_\mu(k)\cA^{-}_\nu(-k)\rangle,
\eeqn
respectively. Here $A^{\pm}_\mu=(A^1_\mu \pm i A^2_\mu)/\sqrt{2}$ and
the Fourier transformed field, $\cA_\mu^i(k)$, is defined as
\beqn
\cA_\mu^i(k) = L^{-4}\,
\sum\nolimits_x e^{-ik_\nu x_\nu-\frac{i}{2}k_\mu}A_\mu^i(x)\,.
\eeqn
The lattice momentum variable is $p_\mu=(2/a)\, \sin (a k_\mu/2)$, where
$k_\mu= 2\pi n_\mu /(aL_\mu)$ with $n_\mu = 0,...,L_\mu-1$, $\mu=1,\dots,4$.
In terms of $p$ the lattice propagator of a free massive scalar particle
in momentum space has a familiar form, $D(p) \propto 1 \slash {(p^2+m^2)}$.
The local gauge condition corresponding to the maximization of Eq.~\eq{eq:coscosfunc}
takes a simple form, $p_\mu \cA_\mu^3 = 0$.

The most general structure of both diagonal and off-diagonal propagators is
\beq
\label{eq:gendmunu}
D_{\mu\nu}(p) = \Bigl(\delta_{\mu\nu} - \frac{p_\mu p_\nu}{p^2}\Bigr)\, D_t(p^2)
+ \frac{p_\mu p_\nu}{p^2}D_l(p^2)\,,
\eeq
where $D_{t,l}$ are the scalar functions. We have three independent
formfactors, $D^{\mathrm{diag}}_t$ and $D_{t,l}^{\mathrm{offdiag}}$, because
in the Landau gauge $D^{diag}_l = 0$.

Below we present the results on $32^4$ lattice at $\beta = 2.40$ which
corresponds to the lattice spacing $a = (1.66 \GeV)^{-1}$.
We generated 30 configurations which are fixed to the MA gauge
by the Simulated Annealing algorithm~\cite{Bali:1996dm} with 10
randomly generated gauge copies. Then the Landau gauge is fixed by the local
over-relaxation algorithm with 20 randomly generated gauge copies.

\begin{figure}[htb]
\vskip -10mm
\centerline{\includegraphics[angle=0,scale=0.27,clip=true]{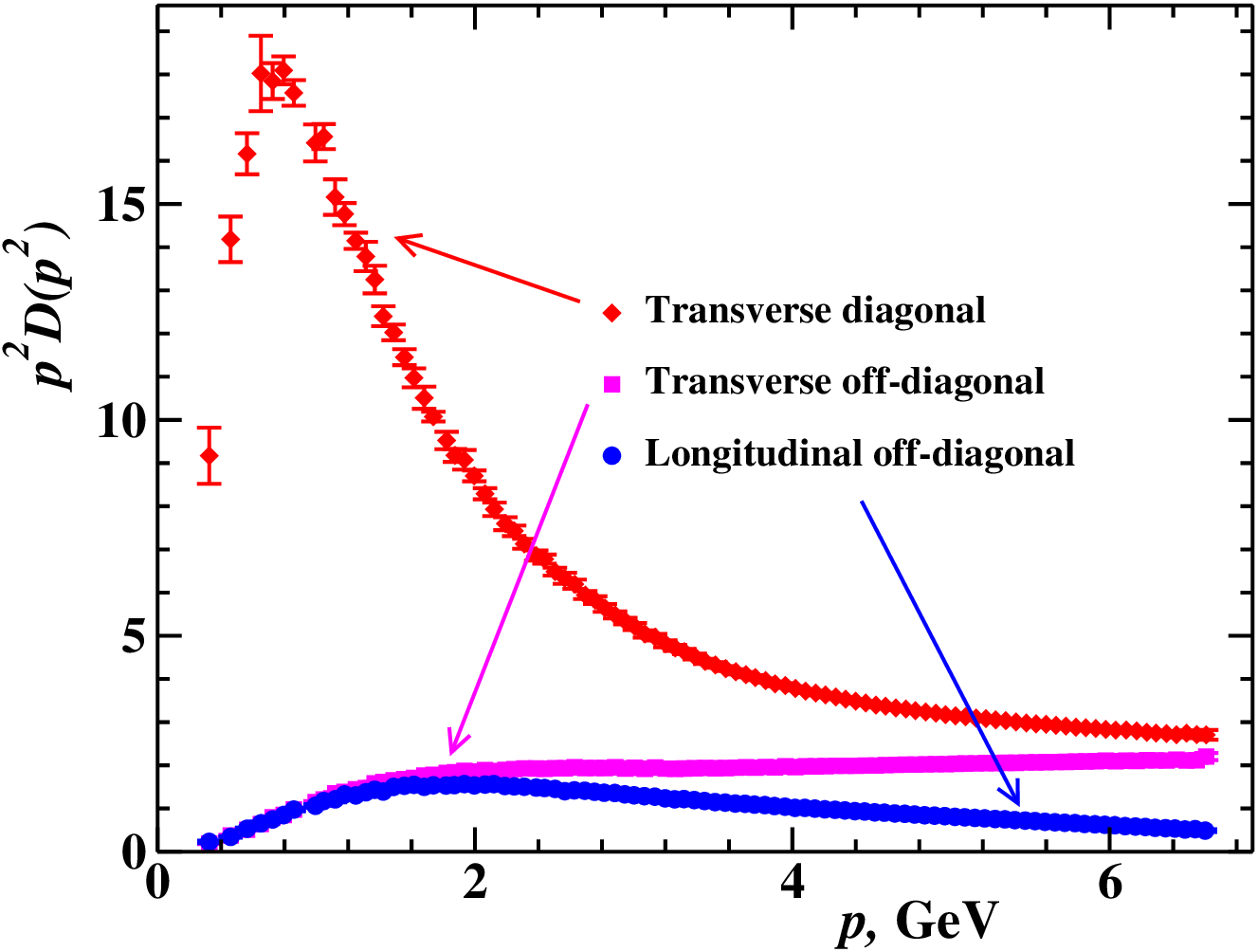}}
\vskip -10mm
\caption{The gluon dressing functions
$p^2\, D(p^2)$.
}
\label{fig:prop}
\end{figure}

We show the gluon dressing functions, $p^2\, D_{t,l}(p^2)$,
{\it vs.} momentum, $p$, in Figure~\ref{fig:prop}. The diagonal (transverse)
dressing function has a relatively narrow maximum at
non--zero momentum $p^{\diag}_0 \approx 0.7 \GeV$. Its behavior
at small momenta is qualitatively very similar to the behavior
of the gluon propagator in the Landau gauge (see, {\it e.g.}, \cite{fit12}). The
longitudinal part of the off-diagonal dressing function has a wide maximum at
$p^{\offdiag}_0 \approx 2 \GeV$. The transverse off-diagonal
dressing function is a monotonically rising function for all available
momenta.

The formfactors $D^{\offdiag}_t(p^2)$ and $D^{\offdiag}_l(p^2)$ coincide
at small momenta. This implies that in the IR region the off--diagonal
propagator is
\beqn
D^{\offdiag}_{\mu\nu}(p) \approx \delta_{\mu\nu} \!\cdot\!
D_t^{\offdiag}(p^2)\,, \,\,\!  p^2 \lesssim 1\GeV\,, \nonumber
\eeqn
\begin{figure}[htb]
\vskip -14mm
\centerline{\includegraphics[angle=0,scale=0.27,clip=true]{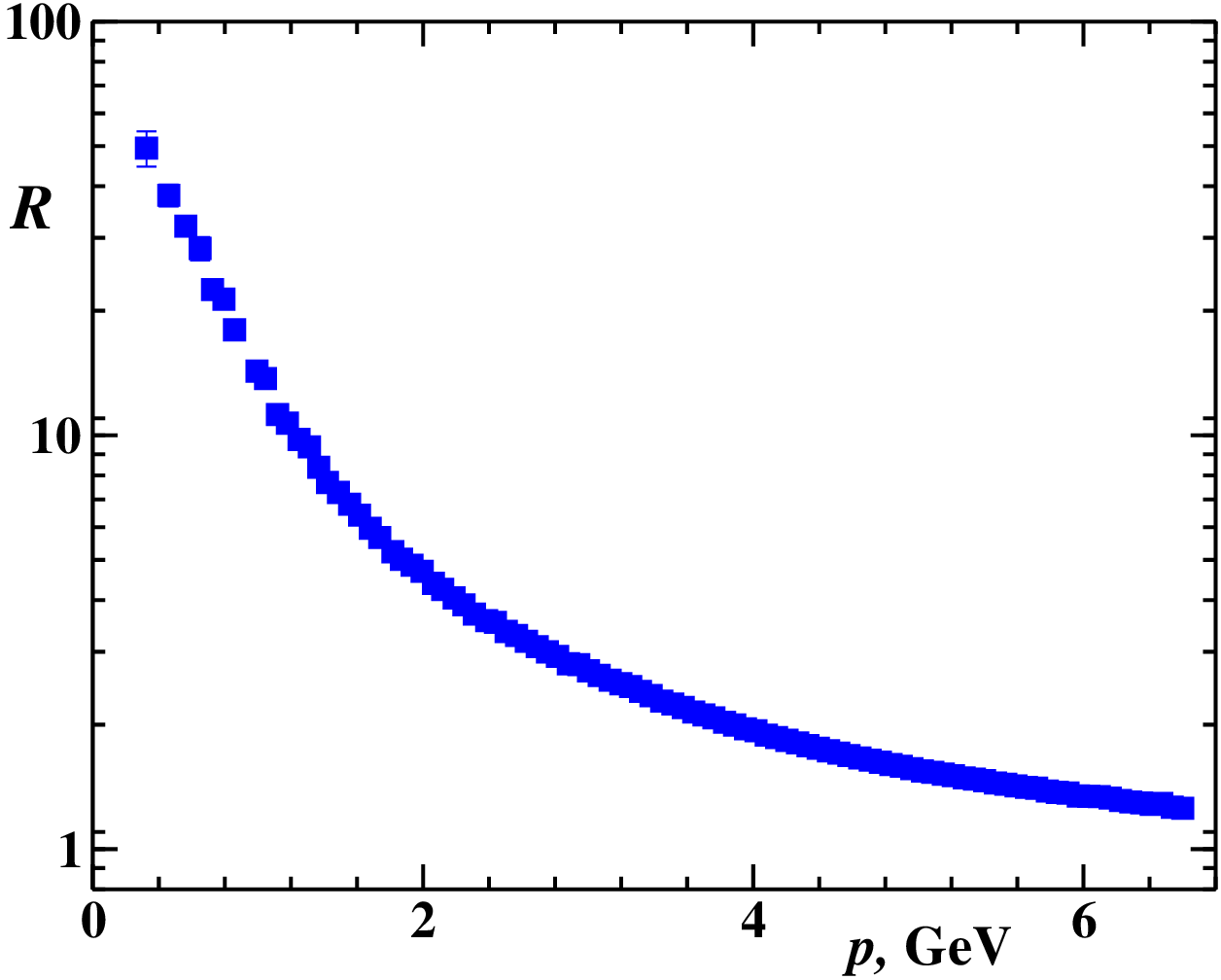}}
\vskip -8mm
\caption{The diagonal/off--diagonal ratio~\eq{eq:ratio:diag}.}
\label{fig:ratio}
\vskip -6mm
\end{figure}

The diagonal formfactor is dominating over the off--diagonal ones.
In Figure~\ref{fig:ratio} we plot the ratio
\beqn
R(p^2) = D^{\diag}_t(p^2) / D^{\offdiag}_t(p^2)\,.
\label{eq:ratio:diag}
\eeqn
which reaches the value $50$ at the smallest available momentum, $p=325$~MeV,
suggesting that the diagonal gluons are responsible for infrared physics.
At higher momenta the suppression of the off-diagonal propagator
becomes weaker.

We fit the formfactors at low momenta by the functions~\cite{ref:itep:propagators,fit12,Chernodub:2001mg}:
\beqn
\label{eq:fit:MIP}
D(p^2) =  &\frac{Z\, m^{2\alpha}}{{(p^2 + m^2)}^{1+\alpha}}\,,
\, &\mbox{(fit 1)}\,, \\
\label{eq:fit:CIS}
D(p^2) =  & \frac{Z\, m^{2\alpha}}{p^{2(1+\alpha)} +
m^{2(1+\alpha)}}\,, \, &
\mbox{(fit 2)}\,,\\
\label{eq:fit:Yukawa}
D(p^2) = &\frac{Z}{p^2 + m^2}\,,\, & \mbox{(Yukawa fit)}\,,\\
\label{eq:fit:Yukawa2}
D(p^2) = & \frac{Z}{m^2 + p^2 + \kappa p^4/m^2}\,,\, &
\mbox{(Yukawa 2 fit)}\,,\\
\label{eq:fit:Gribov}
D(p^2) = & \frac{Z\, p^2}{p^4 + m^4}\,,\, & \mbox{(Gribov fit)}\,,
\eeqn
where $Z$, $\alpha$ , $m$ and $\kappa$ are fitting parameters.
\begin{figure}[!thb]
\vskip -6mm
\centerline{\includegraphics[angle=0,scale=0.27,clip=true]{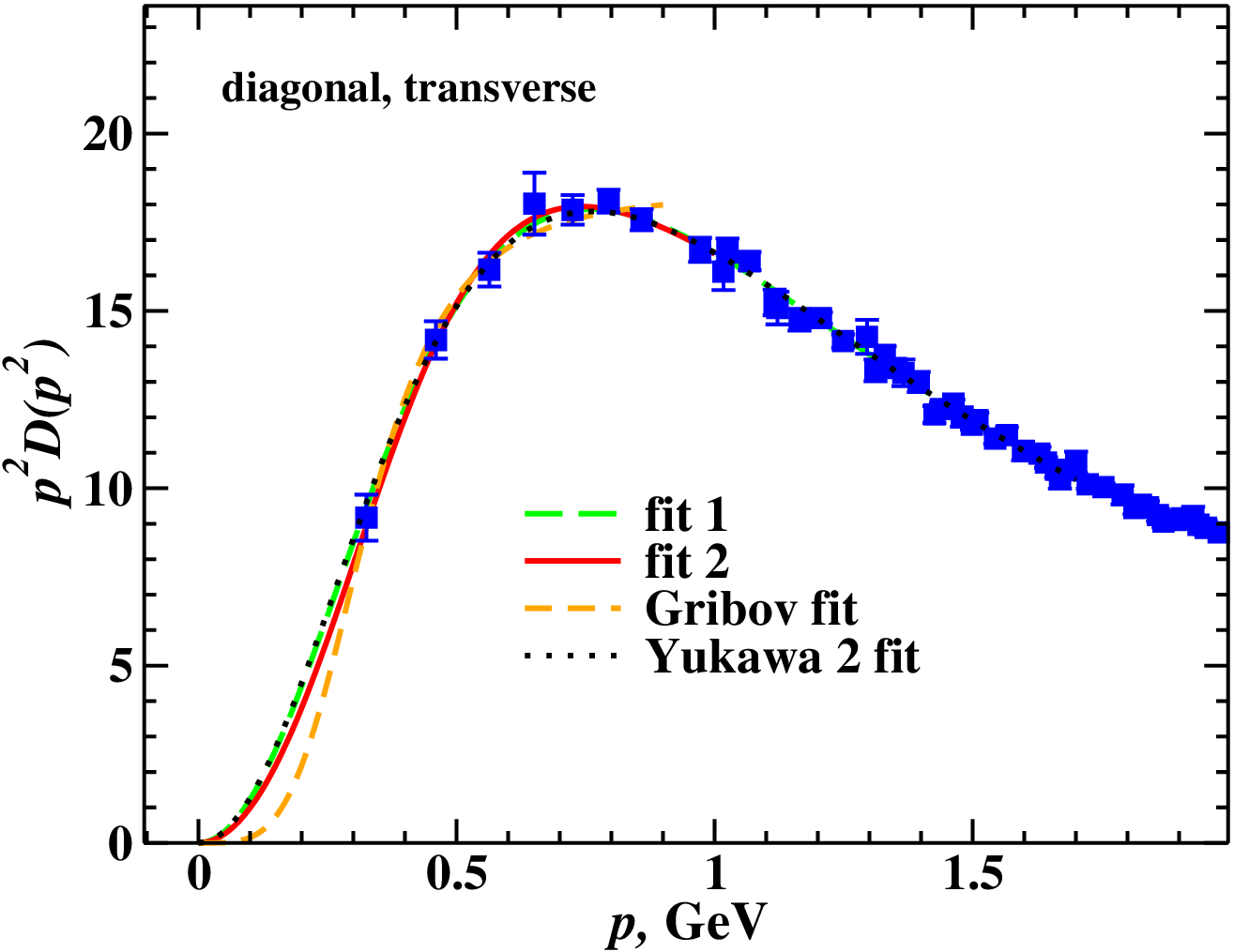}}
\vskip -8mm
\caption{Fits~(\ref{eq:fit:MIP}--\ref{eq:fit:CIS},\ref{eq:fit:Yukawa2}-\ref{eq:fit:Gribov})
of the transverse diagonal formfactor at low momenta.}
\label{fig:fit:prop}
\vskip -5mm
\end{figure}
\begin{table}[!thb]
\vskip -10mm
\begin{center}
\begin{tabular}{|c|c|c|c|c|c|}
\hline
fit  &  $m$, GeV  & $\alpha$ or $\kappa$ & Z \\
\hline
fit 1   & 0.73(2) & 0.92(3) & 16.9(4) \\
fit 2   & 0.58(2) & 0.49(5) & 8.5(2)  \\
Gribov   & 0.33(1) & -  & 4.58(5) \\
Yukawa 2 & 0.50(2) & 0.19(3) & 8.3(3) \\
\hline
\end{tabular}
\end{center}
\caption{The best parameters of the low momentum
fits~(\ref{eq:fit:MIP}--\ref{eq:fit:CIS},\ref{eq:fit:Yukawa2}-\ref{eq:fit:Gribov})
of the diagonal propagator.}
\label{table:fits}
\vskip -8mm
\end{table}

Here we concentrate on the fits of the diagonal propagator, shown in
Figure~\ref{fig:fit:prop}. The best fit parameters are presented in Table~\ref{table:fits}.
The three parameter fits (\ref{eq:fit:MIP},\ref{eq:fit:CIS},\ref{eq:fit:Yukawa2}) are working well
in a low momentum region ($p_{\max} \leqslant 0.8,0.4,0.9$~GeV, respectively).
The corresponding curves are almost indistinguishable from each other.
The mass parameter $m$ do not coincide in these fits, and
the difference between its values is about $30\%$. We have also applied the two-parameter fits
given by Yukawa \eq{eq:fit:Yukawa} and Gribov \eq{eq:fit:Gribov} functions. The Gribov fit is
working well in the region $p_{\max} \leqslant 0.9$~GeV. The Yukawa fit of the diagonal propagator
does not work at all (we get $\chi^2 / d.o.f. \sim 6$ for fits in $p_{\max} < 1 \GeV$ region).

Summarizing, our results obtained in the MA gauge of $SU(2)$
gluodynamics clearly show that in the infrared region the propagator of
the diagonal gluon is much larger (about 50 times at lowest available momentum)
than the propagator of  the off-diagonal gluon. Thus the colored
objects at large distances interact mainly due to an exchange by the diagonal gluons
in agreement with the Abelian dominance~\cite{Ezawa,Suzuki:1989gp}.
The diagonal propagator has qualitatively the same momentum dependence
as the gluon propagator in Landau gauge.
All infrared fits suggest that the diagonal propagator contain massive parameters,
although it is not of the Yukawa form. Further results on the propagators
in the MA gauge can be found in Ref.~\cite{ref:itep:propagators}.

\end{document}